\documentclass[final,3p,times,twocolumn]{elsarticle}

\usepackage{graphicx}

\newcommand{\lw}[1]{\smash{\lower 1.5ex\hbox{#1}}}
\newcommand{\mapright}[1]{\smash{\mathop{\hbox to 1cm{\rightarrowfill}}\limits^{#1}}}
\newcommand{\diff}[2]{\frac{\partial #1}{\partial #2}}
\newcommand{\lsim}{\raisebox{-0.55ex}{$\stackrel{\displaystyle <}{\sim}$}}

\journal{Physics Letters A}

\begin{document}

\begin{frontmatter}

\title{Analysis of residual spectra and the monopole spectrum for 3 K blackbody radiation by means of non-extensive thermostatistics}

\author[1]{Minoru Biyajima}
\address[1]{Department of Physics, Shinshu University, Matsumoto 390-8621, Japan}
\ead{biyajima@azusa.shinshu-u.ac.jp}
\author[2]{Takuya Mizoguchi}
\address[2]{Toba National College of Maritime Technology, Toba 517-8501, Japan\corref{telephone/fax: +81-599-25-8088}}
\ead{mizoguti@toba-cmt.ac.jp}

\begin{abstract}
We analyze residual spectra of 3 K blackbody radiation (CMB) using non-extensive thermostatistics with a parameter $q-1$. The limits of $|q-1|< 1.2\times 10^{-5}$ and the temperature fluctuation $|\delta T|< (1.6\sim 4.3)\times 10^{-5}$ are smaller than those by Tsallis et al. Moreover, analyzing the monopole spectrum by a formula including the chemical potential $\mu$, we obtain the limits $|q-1| < 2.3\times 10^{-5}$ and $|\mu| < 1.6\times 10^{-4}$. $|q-1|$ is comparable with the Sunyaev-Zeldovich effect $y$.
\end{abstract}


\end{frontmatter}

\section{\label{sec1}Introduction}
Very recently, it was reported~\cite{iwasaki2012} that there is a similarity between the blackbody radiation law (i.e., the Planck distribution) and X-ray spectrum. An interesting experimental investigation for this was carried out in~\cite{kawai2012}.The authors cited several papers related to the non-extensive formulas for the Planck distribution~\cite{tsallis1995,plastino1995,tirnakli1998}. More useful information on non-extensive thermostatistics is found in Ref.~\cite{tsallis2009}. The present study is relating to the non-extensive formulas used in the description of the 3 K blackbody radiation (cosmic microwave background (CMB))~\cite{tsallis1995}.

The essence of Ref.~\cite{tsallis1995} is given as follows: The Planck distribution is expressed as 
\begin{eqnarray}
  U_{\rm Planck}(T,\,\nu) = \frac{8\pi h \nu^3}{c^3}\frac{1}{e^x - 1},
  \label{eq1}
\end{eqnarray}
where $x=h\nu/kT$; $h$, $k$ and $T$ are the Planck's constant, the Boltzmann's constant and the temperature, respectively. $c$ is the speed of the light. The non-extensive formula for the Planck distribution is computed as 
\begin{eqnarray}
  &&\hspace*{-12mm} U^{\rm (NETD\:I)}(T,\,\nu,\,q) = U_{\rm Planck}(T,\,\nu)[1-e^{-x}]^{(q-1)}\nonumber\\
  &&\hspace*{-12mm} \quad \times\left\{ 1+(1-q)x\left[\frac{1+e^{-x}}{1-e^{-x}}
  -\frac x2\frac{1+3e^{-x}}{(1-e^{-x})^2} \right] \right\},
  \label{eq2}\\
  &&\hspace*{-12mm} \mapright{\ (q-1)\ll 1\ }\ U_{\rm Planck}(T,\,\nu)
  +\frac{8\pi h\nu^3}{c^3}\frac{q-1}{e^x-1}\nonumber\\
   &&\hspace*{-12mm} \quad \times\left[ \ln(1-e^{-x}) - x\frac{1+e^{-x}}{1-e^{-x}}
  + \frac{x^2}2\frac{1+3e^{-x}}{(1-e^{-x})^2} \right],
  \label{eq3}
\end{eqnarray}
where $(q-1)$ is named the non-extensive or Tsallis parameter. 

\begin{figure}[htbp]
  \begin{center}
  \includegraphics[height=80mm]{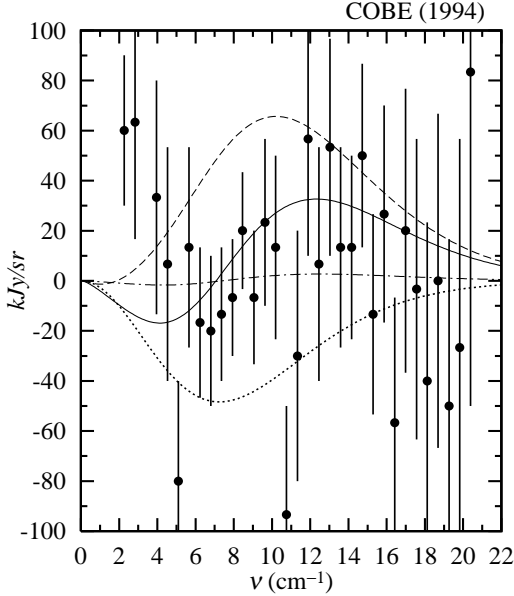}
  \vspace{-3mm}
  \caption{\label{fig1}Analysis of the first residual spectrum (1994). Our confirmation of the analysis by Tsallis et al.~\cite{tsallis1995} (1995). Dashed line: Eq.~(\ref{eq3}) without $U_{\rm Planck}$ using $(q-1) = 3.6\times 10^{-5}$. Dotted line: Eq.~(\ref{eq4}) with $\delta T = -0.1$ mK. Solid line: Sum of them ($\chi^2/NDF = 39/32$). Dashed-dotted line: Ours by using the method of minimum-$\chi^2$; $(q-1) = 3.2\times 10^{-6}$ and $\delta T = -9.4\times 10^{-3}$ mK ($\chi^2/NDF = 32/32$).
}
  \vspace{-3mm}
  \end{center}
\end{figure}

The authors of Ref.~\cite{tsallis1995} analyzed the residual spectrum of the 3 K blackbody radiation (CMB) reported by the NASA COBE Collaboration in 1994~\cite{mather1994}, which can be called the first residual spectrum. The residual spectrum is computed as follows:
\begin{eqnarray*}
  &&\hspace*{-12mm} {\rm [residual\ spectrum]}\nonumber\\ 
  &&\hspace*{-12mm} \quad ={\rm [COBE\ data] - [Eq.\ (\ref{eq1})\ with}\ T=T_{\rm CMB}\ {\rm K]}
\end{eqnarray*}
In their analyses, they utilized Eq.~(\ref{eq3}) with Eq.~(\ref{eq1}), and the following formula for the temperature fluctuation in Eq.~(\ref{eq1}),
\begin{eqnarray}
  \diff{U_{\rm Planck}}{T}\delta T = \frac{e^x x}{(e^x-1)^2}\frac{\delta T}{T},
  \label{eq4}
\end{eqnarray}
where $\delta T$ is the temperature fluctuation (a constant number) and $T=T_{\rm CMB}$. Our confirmation of their analysis and our analysis are shown in Fig.~\ref{fig1}.

Plastino et al.~\cite{plastino1995}, also used Eq.~(\ref{eq3}) to estimate the parameter $(q-1)$ in the analysis of 3 K blackbody radiation.

It is noteworthy mentioning the non-extensive formula in the dilute gas approximation~\cite{tirnakli1998}. The Bose-Einstein distributions in that approximation is computed as,
\begin{eqnarray}
  &&\hspace*{-14mm} U^{\rm (NETD\:II)}(T,\,\nu,\,\mu,\,q)\nonumber\\
  &&\hspace*{-14mm}  = \frac{8\pi h\nu^3}{c^3}\frac{1}{[1+(q-1)(x+\mu)]^{1/(q-1)}-1}\nonumber\\
  &&\hspace*{-14mm} \mapright{\ (q-1)\ll 1\ }\ 
  \frac{8\pi h\nu^3}{c^3}\left[\frac 1{e^x-1} -\frac{\mu e^x}{(e^x-1)^2} + \frac{q-1}2\frac{e^xx^2}{(e^x-1)^2}\right].\nonumber\\
  \label{eq5}
\end{eqnarray}

Moreover, on the basis of the assumption of a g-on gas, Ertik et al~\cite{ertik2009} obtained the following expression with the Mittag-Leffler (ML) function. (See also~\cite{suzuki2002}.)
\begin{eqnarray}
  &&\hspace*{-14mm} U^{\rm (NETD\:III)}(T,\,\nu,\,\mu,\,\alpha) = \frac{8\pi h \nu^3}{c^3}\frac{1}{E_{\alpha}(x+\mu) - 1}\nonumber\\
  &&\hspace*{-14mm} \mapright{\ (\alpha -1)\ll 1\ }\ 
  \frac{8\pi h\nu^3}{c^3}\left[\frac 1{e^x-1} -\frac{\mu e^x}{(e^x-1)^2} + (\alpha -1)\frac{f(x)}{(e^x-1)^2}\right],\nonumber\\
  \label{eq6}
\end{eqnarray}
where
\begin{eqnarray*}
E_{\alpha}(x+\mu) &\!\!\!=&\!\!\! \sum_{n=0}^{\infty} \frac{(x+\mu)^n}{\Gamma(n\alpha + 1)},\\
f(x) &\!\!\!=&\!\!\! \sum_{k=0}^{\infty} \frac{kx^k\psi(1+k)}{\Gamma(1+k)}, 
\end{eqnarray*}
$\psi(z) = d(\ln \Gamma (z))/dz$ is the digamma function, and $(\alpha -1)$ is named the fractional parameter.

Several estimated values by means of non-extensive thermostatistics are summarized in Table~\ref{tab1}.
\begin{table*}[htbp]
  \caption{\label{tab1}Values and limits estimated by using non-extensive thermostatistics.}
  \vspace{-2mm}
  \begin{center}
  \renewcommand{\arraystretch}{1.2}
  \begin{tabular}{c|c|l|l}
  \hline
  Authors & Formula & data & Estimated limits, $\chi^2/NDF$\\
  \hline
  Tsallis et al~\cite{tsallis1995} & $U^{\rm (NETD\:I)} - U_{\rm Planck} $ & residual & $(q-1)=3.6\times 10^{-5}$ (fixed),\\ 
  & $+ (\partial U_{\rm Planck}/\partial T)\delta T$ &  spectrum & $\delta T =-1.0\times 10^{-4}$, 39.3/32\\
  \cline{1-1}\cline{4-4}
  Our analysis & (minimum-$\chi^2$) & (RS) (1994) & $(q-1)=(0.3\pm 1.4)\times 10^{-5}$,\\
  & & & $\delta T=(-0.9\pm 5.1)\times 10^{-5}$, 32.2/32\\
  \hline
  Plastino et al~\cite{plastino1995} & $U^{\rm (NETD\:I)} - U_{\rm Planck}$ & RS (1994) & $|q-1|<5.3\times 10^{-4}$\\
  \hline
  Tirnakli et al~\cite{tirnakli1998} & $U^{\rm (NETD\:II)}$ & RS (1994) & $|q-1|<4.1\times 10^{-5}$\\
  \hline
  Ertik et al~\cite{ertik2009} & $U^{\rm (NETD\:III)}$ & monopole & $T_{\rm CMB}=2.72842$ K,\\ 
  $\alpha$: fractional para. && spec. (1996) & $\alpha -1 = -1.98\times 10^{-5}$ \\
  \hline
  \end{tabular}
  \end{center}
  \vspace{-3mm}
\end{table*}

On the other hand, the NASA COBE Collaboration published the residual spectrum computed from the full COBE data in 1996~\cite{fixen1996}. Hereafter we call it the second residual spectrum. Of course, the second one is different from that in 1994~\cite{mather1994}. Moreover, the Collaboration released their data on monopole spectrum of CMB on their web site in 2005. We summarize the various sources of the spectra reported by the NASA COBE Collaboration~\cite{mather1994,fixen1996,fixsen2002,nasa2005} in Table~\ref{tab2}. (See also Refs.~\cite{henry1968,kompaneets1957,weymann1965,zeldovich1969}.)

Thus, our purpose of this paper is to analyze the second residual spectrum reported in 1996 and the monopole spectrum released in 2005, by means of Eqs.~(\ref{eq1}), (\ref{eq3}), (\ref{eq4}), (\ref{eq5}) and (\ref{eq6}).

This paper is divided in to following paragraphs. In the second paragraph, we summarize data sources by the NASA COBE Collaboration with an explanation on the chemical potential $\mu$ and the Sunyaev-Zeldovich (S-Z) effect $y$~\cite{kompaneets1957,weymann1965,zeldovich1969}. In the third paragraph, we analyze the monopole spectrum~\cite{nasa2005} as well as residual spectrum by means of non-extensive thermostatistics.

\section{\label{sec2}Data sources by NASA COBE Coll. -- Including chemical potential $\mu$ and Sunyaev-Zeldovich effect $y$ --}
We should mention the roles of the dimensionless chemical potential $\mu$ and of the Sunyaev-Zeldovich effect $y$~\cite{kompaneets1957,weymann1965,zeldovich1969,sugiyama2001,durrer2008} introduced in Table~\ref{tab2}. The former describes the distortion of the Planck distribution for the cosmic microwave background. The latter is reflecting the influence due to scattering of hot electrons with the CMB photons.
\begin{table*}[htbp]
  \caption{\label{tab2}Various data sources reported by the NASA COBE Collaboration. Cosmic microwave background temperature $T_{\rm CMB}$, limits of chemical potential $|\mu|$ and the S-Z effect $|y|$ are presented by the NASA COBE Collaboration. Notice that the dipole spectrum is an observed quantity, which is described by an expression with $T_{amp}$, $(T_{amp}/T)(xe^x/(e^x-1)^2)$~\cite{henry1968}.}
  \vspace{-2mm}
  \begin{center}
  \renewcommand{\arraystretch}{1.2}
  \begin{tabular}{c|l|l}
  \hline
  Year & Data & Refs.\quad Estimated values\\
  \hline
  1994 & Partial COBE data & \cite{mather1994} ApJ. 420, 439-444 (1994)\\
       & Residual spectrum (numerical values) & $T_{\rm CMB} = 2.726\pm 0.010$ K (95\% CL)\\
  \hline
  1996 & Full COBE data & \cite{fixen1996} ApJ. 476, 576-587 (1996)\\
       & $\cdot$ Dipole spectrum (numerical values) & $T_{\rm amp}=3.372$ mK, and $T_{dipole}=2.717$ K\\
       & $\cdot$ Monopole spectrum (Fig. only) & $T_{\rm CMB} = 2.728\pm 0.004$ K\\
       & $\cdot$ Residual spectrum (numerical values) & $|\mu|<9\times 10^{-5}$, $|y|<1.5\times 10^{-5}$ (95\% CL)\\
  \hline
  2002 & Summary of estimated values & \cite{fixsen2002} ApJ. 581, 817-822 (2002)\\
       & ($|\mu|<9\times 10^{-5}$, $|y|<1.5\times 10^{-5}$ (95\% CL)) & $T_{\rm CMB} = 2.725\pm 0.001$ K\\
  \hline
  2005 & Monopole spectrum (numerical values) & \cite{nasa2005} NASA COBE Web site\\
       & ($T_{\rm CMB} = 2.725\pm 0.001$ K) & http://lambda.gsfc.nasa.gov/product\\
       & & /cobe/firas\_monopole\_get.cfm\\
  \hline
  \hline
  cf.  & Our analysis of monopole spectrum~\cite{nasa2005} & $T_{\rm CMB} = 2.725\pm 4\times 10^{-5}$ K,\\
       &  by Eq.~(\ref{eq10}). & $|\mu|<1.4\times 10^{-4}$, $|y|<1.1\times 10^{-5}$\\
  \hline
  \end{tabular}
  \end{center}
  \vspace{-3mm}
\end{table*}

The dimensionless chemical potential $\mu$ introduced in the Planck distribution Eq.~(\ref{eq1}) is necessary in describing the Compton scattering $\gamma + e^- \rightleftharpoons \gamma + e^-$ in an early Universe, because the number of photons is conserved in that scattering.  
\begin{eqnarray}
  U_{\rm BE}(T,\,\nu,\,\mu) &\!\!\!=&\!\!\! \frac{8\pi h \nu^3}{c^3}\frac{1}{e^{x+\mu} - 1}\nonumber\\
  &\!\!\! \approx &\!\!\! \frac{8\pi h \nu^3}{c^3}\left[\frac 1{e^{x}-1} - \mu\frac{e^x}{(e^{x}-1)^2}\right],
  \label{eq7}
\end{eqnarray}
where BE stands for the Bose-Einstein distribution. This calculation is also performed in Eqs.~(\ref{eq5}) and (\ref{eq6}).

The Sunyaev-Zeldovich (S-Z) effect is given as
\begin{eqnarray}
  {\rm S-Z\ effect}\ =\ 
   \frac{C_B\nu^3 yxe^x}{(e^x-1)^2}\left(x\coth \frac x2-4\right)
  \label{eq8},
\end{eqnarray}
where $C_B = 8\pi h/c^3$. Moreover, $y$ is the parameter for the inverse Compton scattering,
\begin{eqnarray}
  y = \int dl n_e \sigma_T \frac{kT_e}{m_ec^2}
  \label{eq9},
\end{eqnarray}
where $l$, $n_e$, $\sigma_T$ and $T_e$ are the size of the high-temperature region in the Universe, the number density of electrons, the cross section of Thomson scattering and temperature of electron, respectively. See footnote~\footnote{\label{foot1}
Notice that the S-Z effect is related to the temperature fluctuation $\Delta T(x)$ (a function of $x$) through the Kompaneets equation~\cite{kompaneets1957,durrer2008}:
\begin{eqnarray*}
  \delta u = yx^{-2} \diff{}{x}\left(x^4\diff{u}{x}\right)
\end{eqnarray*}
where $U_{\rm Planck}/(8\pi h\nu^3/c^3) = u$.
\begin{eqnarray*}
  \frac{\Delta T(x)}T = \frac uT\frac{\delta u}u \left(\frac{du}{dT}\right)^{-1} 
  = \frac{e^x-1}{xe^x}\frac{\delta u}u
  = -y\left[4-x\coth\frac x2\right].
\end{eqnarray*}
As we use $U^{\rm (NETD\:II)}/(8\pi h\nu^3/c^3)$ (Eq.~(\ref{eq5})), we obtain a more complicated expression as follows:
\begin{eqnarray*}
  &&\hspace*{-14mm}  \frac{\Delta T(x)}T = -y\left[4-x\coth\frac x2\right] 
  -y\mu\left[(4-x)\coth\frac x2-x\frac{1+3\coth\frac x2}{e^x-1}\right] \\
  &&\hspace*{-10mm} -\frac{q-1}2 yx\left[8-(5x^2+8x-2)\coth\frac x2 + \frac{4(x^2-1)}{e^x-1} + 6\left(\frac{e^xx}{e^x-1}\right)^2\right].
\end{eqnarray*}
}.

Their limits ($|\mu|$ and $|y|$) presented in Table~\ref{tab2} are estimated in terms of the following formula for the description of the monopole spectrum:
\begin{eqnarray}
\hspace*{-6mm}  U^{\rm (NASA\ COBE)} \approx U_{\rm BE} 
   + \frac{C_B\nu^3 yxe^x}{(e^x-1)^2}\left(x\coth \frac x2-4\right).
  \label{eq10}
\end{eqnarray}
From the analysis of the monopole spectrum with Eq.~(\ref{eq10}), we obtain the following values: $T = 2.7250\pm 4\times 10^{-5}$ K, $\mu = (-2.6\pm 5.6)\times 10^{-5}$ and $y = (1.6\pm 4.8)\times 10^{-6}$ ($\chi^2/NDF = 44.9/41$), i.e., $|\mu| < 1.4\times 10^{-4}$ and $|y| < 1.1\times 10^{-5}$ (95\% CL).

\begin{figure}[htbp]
  \begin{center}
  \includegraphics[height=80mm]{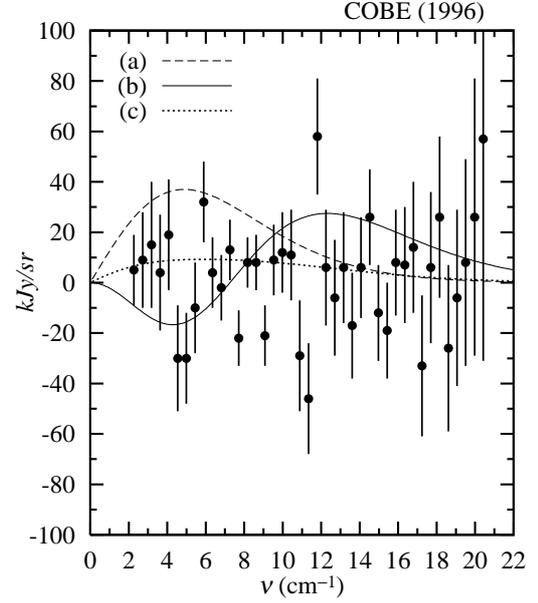}
  \vspace{-3mm}
  \caption{\label{fig2}Confirmation of NASA CMB parameters: (a) $\mu = -9\times 10^{-5}$ ($\chi^2/NDF = 78/42$). (b) $y=1.5\times 10^{-5}$ ($\chi^2/NDF = 120/42$). (c) Our calculation by Eq.~(\ref{eq10}) with $\mu =-2.6\times 10^{-5}$ and $y=1.6\times 10^{-6}$ ($\chi^2/NDF = 51.7/41$).}
  \vspace{-3mm}
  \end{center}
\end{figure}

The second residual spectrum on the 3 K blackbody radiation (CMB) is shown in Fig.~\ref{fig2}. The utilized values are taken from those in Table~\ref{tab2}. 

\section{\label{sec3}Analyses of monopole and residual COBE spectra by means of non-extensive thermostatistics}
\begin{figure}[htbp]
  \begin{center}
  \includegraphics[height=62mm]{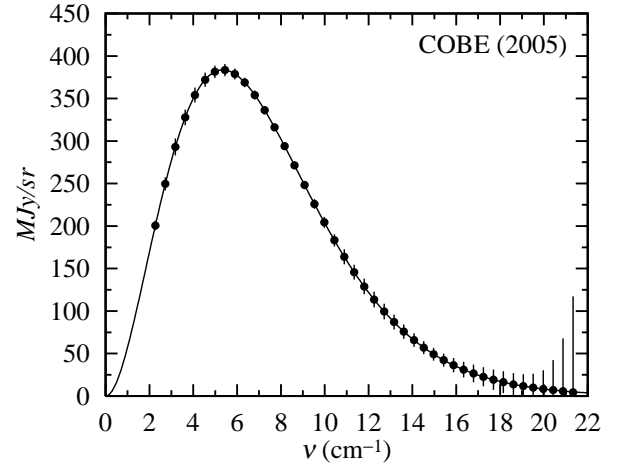}
  \vspace{-3mm}
  \caption{\label{fig3}Analysis of monopole spectrum of CMB by means of Eq.~(\ref{eq3}): $T=2.725$ K, $(q-1) = 1.2\times 10^{-5}$ and $\chi^2/NDF = 48.9/41$. Error bars are 400$\sigma$.}
  \vspace{-3mm}
  \end{center}
\end{figure}

\begin{figure}[htbp]
  \begin{center}
  \includegraphics[height=80mm]{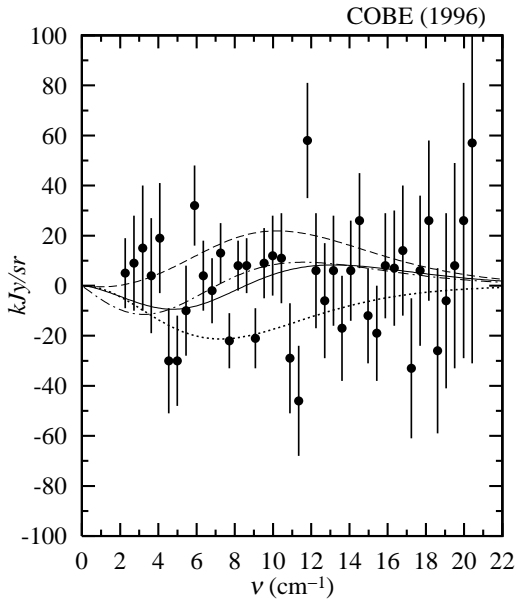}
  \vspace{-3mm}
  \caption{\label{fig4}Analysis of the 2nd residual spectrum (1996)~\cite{fixen1996}. Dashed line: Eq.~(\ref{eq3}) with $(q-1) = 1.2\times 10^{-5}$. Dotted line: Eq.~(\ref{eq4}) with $\delta T = -4.3\times 10^{-5}$ K. Solid line: Sum of them. $\chi^2/NDF = 49.6/41$. Dashed-dotted-dashed line: From limits by Eq.~(\ref{eq5}), $\mu=0.4\times 10^{-4}$ and $(q-1) = 0.58\times 10^{-5}$ in Table~\ref{tab4} are chosen. $\chi^2/NDF = 51.6/41$.}
  \vspace{-3mm}
  \end{center}
\end{figure}

Since the monopole spectrum of CMB is presented in~\cite{nasa2005}, we are able to analyze it by means of Eqs.~(\ref{eq1})--(\ref{eq3}). In the present analysis, the temperature can be determined with the CERN MINUIT program. Our results are shown in Figs.~\ref{fig3} and \ref{fig4} and Table~\ref{tab3}. 
\begin{table*}[htbp]
  \caption{\label{tab3}Analysis of monopole spectrum (2005) and residual spectrum (1996) by means of $U^{\rm (NETD\:I)}$~\cite{tsallis1995}. Notice that the same errors appear in the second column, because of a sign of ``failure'' in the CERN MINUIT program.}
  \vspace{-5mm}
  \begin{center}
  \renewcommand{\arraystretch}{1.2}
  \begin{tabular}{lcccc}
  \hline
  Formulas & $T$ (K) & $(q-1)$ & $\delta T$ (K) & $\chi^2$/NDF\\
  \hline
  $U^{\rm (NETD\:I)}$ & $2.7250\:(\pm 2\times 10^{-5})$ & $(-0.5\pm 5.9)\times 10^{-6}$ & --- & 45.1/41\\
   &  & $<1.23\times 10^{-5}$ & & (95 \% CL)\\
  \hline
  $U_{\rm Planck} + \frac{e^x x}{(e^x-1)^2}\frac{\delta T}{T}$ & $2.7250\pm 6.8\times 10^{-4}$ & --- & $8.1\times 10^{-6}\pm 6.8\times 10^{-4}$ & (failure)\\
  \hline
  \hline
  $U^{\rm (NETD\:I)}-U_{\rm Planck} + \frac{e^x x}{(e^x-1)^2}\frac{\delta T}{T}$  & 2.7250 (fixed) & $(-0.2\pm 5.9)\times 10^{-6}$ & $(0.1\pm 2.1)\times 10^{-5}$ & 45.0/41\\
  &  & $<1.2\times 10^{-5}$ & $<4.3\times 10^{-5}$ & (95 \% CL)\\
  \hline
  $U^{\rm (NETD\:I)}-U_{\rm Planck}$ & 2.7250 (fixed) & $(0.6\pm 20.9)\times 10^{-7}$ & --- & 45.0/42\\
  &  & $<4.2\times 10^{-6}$ &  & (95 \% CL)\\
  \hline
  $\frac{e^x x}{(e^x-1)^2}\frac{\delta T}{T}$ & 2.7250 (fixed) & --- & $(0.3\pm 7.6)\times 10^{-6}$ & 45.0/42\\
  &  &  & $<1.6\times 10^{-5}$ & (95 \% CL)\\
  \hline
  \end{tabular}
  \end{center}
  \vspace{-3mm}
\end{table*}

It is found that in the analysis of the monopole spectrum of CMB by Eqs.~(\ref{eq1})--(\ref{eq4}) that the temperature $T$ and $\delta T$ in Eq.~(\ref{eq4}) cannot be determined at the same time. (See the second column in Table~\ref{tab3}.) On the other hand, the non-extensive parameter $(q-1)$ and temperature $T$ are determined simultaneously.

The shape of the solid line (the sum of two magnitudes attributed to $(q-1)$ and $\delta T$) in Fig.~\ref{fig4} can be compared with that of the S-Z effect in Fig.~\ref{fig2}.

Moreover, we analyze the monopole spectrum by means of Eqs.~(\ref{eq5}) and (\ref{eq6}). In those cases, the CMB temperature $T_{\rm CMB}$, the chemical potential $\mu$, the non-extensive parameter $(q-1)$ and/or the fractional parameter $(\alpha -1)$ are all determined simultaneously. Our results are summarized in Table~\ref{tab4}.
\begin{table*}[htbp]
  \caption{\label{tab4}Analysis of the monopole spectrum by $U^{\rm (NETD\:II)}$ (Eq.~(\ref{eq5})) and $U^{\rm (NETD\:III)}$ (Eq.~(\ref{eq6})).}
  \vspace{-2mm}
  \begin{center}
  \begin{tabular}{c|cccc}
  \hline
  & $T$ (K) & $\mu$ & $(q-1)$ or $(\alpha-1)$ & $\chi^2$/NDF\\
  \hline
  \lw{$U^{\rm (NETD\:II)}$}  & $2.72502\pm 3\times 10^{-5}$ & --- & $(-0.53\pm 4.98)\times 10^{-6}$ & 45.1/41\\
  \lw{$(q-1)$} &&& $<1.0\times 10^{-5}$ & (95 \% CL)\\
    \cline{2-5}
  & $2.7250\pm 1\times 10^{-4}$ & $(-3.0\pm 6.3)\times 10^{-5}$ & $(3.4\pm 9.7)\times 10^{-6}$ & 44.9/40\\
  && $<1.6\times 10^{-4}$ & $<2.3\times 10^{-5}$ & (95 \% CL)\\
  \hline
  \lw{$U^{\rm (NETD\:III)}$} & $2.7250\pm 1\times 10^{-4}$ & --- & $(-0.22\pm 1.4)\times 10^{-5}$ & 45.1/41\\
  \lw{$(\alpha-1)$} &&& $<3.0\times 10^{-5}$ & (95 \% CL)\\
    \cline{2-5}
  & $2.7249\pm 2\times 10^{-4}$ & $(-4.4\pm 8.5)\times 10^{-5}$ & $(1.5\pm 3.6)\times 10^{-5}$ & 44.8/40\\
  && $<2.1\times 10^{-4}$ & $<8.7\times 10^{-5}$ & (95 \% CL)\\
  \hline
  \end{tabular}
  \end{center}
  \vspace{-3mm}
\end{table*}

Finally, it should be stressed that, when an additional factor, i.e., the temperature fluctuation Eq.~(\ref{eq4}) is included in Eq.~(\ref{eq5}), the four parameters cannot be determined as far as the CERN MINUIT program is utilized.

\section{\label{sec4}Concluding remarks and discussions}
From the above mentioned studies, we summarize several remarks (R1--R3) and, moreover add to the discussions (D1--D4).\medskip

{\bf R1) } First of all, we analyze the second residual spectrum of 3 K blackbody radiation (CMB)~\cite{fixen1996} by means of Eq.~(\ref{eq3}) ($U^{\rm (NETD\:I)}$)~\cite{tsallis1995}. For the non-extensive parameter $(q-1)$, we obtain $|q-1|< 1.2\times 10^{-5}\sim 4.2\times 10^{-6}$, which is somewhat smaller than that in Table~\ref{tab1} (Ref.~\cite{tsallis1995}).\medskip

Concerning the temperature fluctuation $|\delta T|$ in Table~\ref{tab3}, we obtain the limit $4.3\times 10^{-5}\sim 1.6\times10^{-5}$ which is approximately 50\% smaller than $|\delta T|=0.1$ mK (fixed) in the analysis of the first residual spectrum (1994) in Table~\ref{tab1} (Ref.~\cite{tsallis1995}).\medskip

The non-extensive parameter $|q-1| \lsim 1.2\times 10^{-5}$ and the temperature fluctuation $|\delta T/T|\sim 2.0\times 10^{-5}$ are almost of the same order.\medskip

Moreover, it can be said that a role of the sum of the non-extensive parameter $(q-1)$ and the temperature fluctuation $\delta T$ seems to be an effective S-Z effect, because $|q-1|\sim |y|$. (Compare Fig.~\ref{fig4} with Figs.~\ref{fig2} and \ref{fig1}.)\footnote{\label{foot2}
The following analytic calculation with Eqs.~(\ref{eq4}) and (\ref{eq5}) (instead of Eq.~(\ref{eq3})) may be possible:
\begin{eqnarray*}
  \hspace*{-7mm} \frac{e^x x}{(e^x-1)^2}\frac{\delta T}{T} + \frac{q-1}2\frac{e^xx^2}{(e^x-1)^2} &\!\!\!\!\!=&\!\!\!\!\! \frac{q-1}2\frac{e^xx}{(e^x-1)^2}\left[x+\left(\frac 2{q-1}\right)\frac{\delta T}{T}\right],\\
  &\!\!\!\!\!\approx&\!\!\!\!\! [{\rm an\ effective\ (S-Z)\ effect}].
\end{eqnarray*}
as $\delta T$ is assigned to be negative.
}\medskip

{\bf R2) } When we simultaneously take into account of the dimensionless chemical potential $\mu$~\cite{mather1994,fixen1996,zeldovich1969,sugiyama2001,durrer2008} and the non-extensive parameter $|q-1|$~\cite{tsallis1995,plastino1995,tirnakli1998} in Eq.~(\ref{eq5}) ($U^{\rm (NETD\:II)}$), we obtain the following inequality,
\begin{eqnarray*}
  |\mu| > |q-1|.
\end{eqnarray*}
The limit $|\mu|<1.6\times 10^{-4}$ is almost of the same order as that of Refs.~\cite{fixen1996,fixsen2002}. \medskip

{\bf R3) } The role and magnitude of the fractional parameter $(\alpha -1)$ in Eq.~(\ref{eq6}) ($U^{\rm (NETD\:III)}$) are similar to those of the non-extensive parameter $(q-1)$ in Eq.~(\ref{eq5}). This is seen in Table~\ref{tab4}. However, the magnitude of $(\alpha-1)$ is three times larger than that of $(q-1)$.\medskip

{\bf D1) } It is worthwhile examining the normalization of the Planck distribution in Eq.~(\ref{eq1})~\cite{tsallis1995}: Provided that the magnitude of the distortion in the space-time is finite, we can examine it in the monopole spectrum according to Ref.~\cite{tsallis1995}
\begin{eqnarray}
  U(T,\: \nu,\: \varepsilon) = \frac{\pi^{d/2}(d-1)dh\nu^d}{\Gamma(d/2+1)c^d(e^x-1)},
  \label{eq11}
\end{eqnarray}
where $d=3+\varepsilon$. Our estimated limits are: $T=2.7250\pm 5.0\times 10^{-5}$ and $\varepsilon =(0.3\pm 2.4)\times 10^{-5}$ ($\chi^2/NDF=45.1/41$), i.e., $|\varepsilon|<5.1\times 10^{-5}$ (95 \% CL). The limit of $|\varepsilon|<5.1\times 10^{-5}$ is almost the same as that of $|\delta T|<4.3\times 10^{-5}$ in Table~\ref{tab3}. \medskip

{\bf D2) } In the future, we will elucidate whether or not the non-extensive thermostatistics can describe the CMB spectrum through the quantity $\Delta T(x)/T$~\cite{durrer2008}. (See footnote~\ref{foot1}.)\medskip

{\bf D3) } Third we would like to mention the study by Zeng et al.~\cite{zeng2012}, in which they calculated the following formula on the basis of the assumption of a Kerr Nonlinear Blackbody (KNB),
\begin{eqnarray}
  &&\hspace*{-12mm} U^{\rm (KNB)}(T,\,\nu,\,q)
  \mapright{\ (q-1)\ll 1\ }
  \frac{8\pi h\nu^3 v (T)}{c^4}\nonumber\\
  &&\hspace*{-12mm} \times \left[\frac{1}{e^{x_c}-1} +(q-1)\frac{x_ce^{x_c}(2-2e^{x_c}+3x_c+x_ce^{x_c})}{(e^{x_c}-1)^3}\right],\nonumber\\
  \label{eq12}
\end{eqnarray}
where $x_c$ is a function of $x' = x/(1+\delta T/T)$, $(q-1)$ and $\gamma = v (T)/c$, An explicit expression is given by Eq.~(36) in Ref.~\cite{zeng2012}. See our preliminary analysis~\footnote{\label{foot3}
A preliminary analysis of the monopole spectrum by Eq.~(\ref{eq12}) is made by the present authors: $|q-1|<1\times 10^{-5}$ , $T = 2.7250\pm 3\times 10^{-4}$ K and $|\delta T|<6\times 10^{-4}$ K, provided that $\gamma = 1.0000\pm 0.0001$. The CERN MINUIT program cannot precisely determine them, because the number of parameters is large (4).
}.\medskip

{\bf D4) } The present study is expected to be applicable to other fields, for example, in analyses of the transverse distributions in heavy-ion collisions~\cite{biyajima2005,adare2010}. Indeed, the determination of the temperatures and the non-extensive parameter $(q-1)$ is also important in describing Pb--Pb collisions at LHC energies~\cite{mizoguchi2012}, because the events including a little Big Bang in high energy heavy-ion collisions are expected.

\section*{Acknowledgements}
Authors would sincerely like to thank Prof. J. C. Mather and Prof. D. J. Fixsen for their kindness in communications for NASA COBE data and the method of analysis at an early stage of this investigation. We are also indebted to Dr. J. Kawai for discussion at NEXT2012 held in Nara, Japan. One of the authors (M. B.) would like to thank Prof. N. Sugiyama for insightful conversations.

\bibliographystyle{elsarticle-num}
\bibliography{<your-bib-database>}

\end{document}